\documentstyle[epsfig]{mn2e}

\title[57 second oscillations in Nova Centauri 1986 (V842 Cen)]
{57 second oscillations in Nova Centauri 1986 (V842 Cen)}
\author[Patrick A. Woudt et al.]
       {Patrick A.~Woudt$^1$\thanks{email: Patrick.Woudt@uct.ac.za},
        Brian Warner$^1$, Julian Osborne$^2$, Kim Page$^2$\\
        $^1$ Department of Astronomy, University of Cape Town, Private Bag X3,
        Rondebosch 7701, South Africa\\
        $^2$ Department of Physics and Astronomy, University of Leicester, University
        Road, Leicester LE1 7RH, United Kingdom}
\date{2008 December 9}

\begin{document}

\maketitle

\begin{abstract}
High speed photometry in 2008 shows that the light curve of V842 Cen possesses a 
coherent modulation at 56.825 s, with sidebands at 56.598 s and 57.054 s. These 
have appeared since this nova remnant was observed in 2000 and 2002. We deduce 
that the dominant signal is the rotation period of the white dwarf primary and 
the sidebands are cause by reprocessing from a surface moving with an orbital 
period of 3.94 h. Thus V842 Cen is an intermediate polar (IP) of the DQ Herculis 
subclass, is the fastest rotating white dwarf among the IPs and is the third fastest 
known in a cataclysmic variable. As in other IPs we see no dwarf nova oscillations, 
but there are often quasi-periodic oscillations in the range 350 -- 1500 s.
   There is a strong brightness modulation with a period of 3.78 h, which we 
attribute to negative superhumps, and there is an even stronger signal at 2.886 h 
which is of unknown origin but is probably a further example of that seen in GW Lib 
and some other systems.
  We used the Swift satellite to observe V842 Cen in the ultra-violet and in X-rays, 
although no periodic modulation was detected in the short observations. The X-ray luminosity 
of this object appears to be much lower than that of other IPs in which the accretion region 
is directly visible.

\end{abstract}


\section{Introduction}

Nova Centauri 1986 (later designated V842 Cen) was discovered on 22 November 
1986 at $V = 5.6$ and two days later reached maximum at $V = 4.6$ (McNaught 1986). It 
was a moderately fast nova, with decay time $t_3 = 48$ d and developed an obscuring 
dust shell starting 37 d after maximum and reaching greatest optical thickness 74 d 
after maximum, placing it in the group II category of nova light curves as defined 
by Duerbeck (1981), which is essentially of DQ Herculis type. The pre-eruption 
brightness was estimated by McNaught (1986) to be in the range $B \sim 18.0 - 18.6$.  
Fifteen years after eruption it was at $V \sim 15.8$ (Downes \& Duerbeck 2000; 
Woudt \& Warner 2003), and our latest measurements give $V \sim 16.3$. Therefore 22 years 
after maximum it is still about two magnitudes above its pre-nova brightness, 
which could be a result of irradiation enhanced mass transfer caused by a still 
very hot white dwarf primary, especially if it is relatively massive (see Warner 
(2002) for a discussion of anomalous post-nova luminosities arising from this 
effect), though Kato (2008) reports that there is nothing in the eruptive behaviour 
to suggest a mass much greater than 0.7 M$_{\odot}$. DQ Her, also a moderately fast nova 
($t_3 = 94$ d), but of low mass (0.60 M$_{\odot}$: Horne, Welsh \& Wade 1993), reached its 
maximum optical thickness dust obscured phase 101 d after maximum light; it was 
at $m_{pg} \sim 14.8$ prior to its 1934 eruption (Robinson 1975) and yet had returned 
to that level less than 20 years later (Walker 1956).

    Sekiguchi et al.~(1989) give distance estimates that average 1.0 kpc 
from strengths of interstellar Na D lines and the 2200 {\AA} feature, and a reddening 
of E(B-V) = 0.55. Gill \& O'Brien (1998) in 1995 detected an ejecta shell of 
diameter $\sim 1.5$ arcsec in direct imaging.

   There is nothing in the nova development of V842 Cen that marks it as in any 
way peculiar. The UV and soft X-Ray turn-off times are normal (Gonz\'{a}lez-Riestra, 
Orio \& Gallagher 1998) and the abundances (including a high carbon content typical 
of dusty ejecta) are within the normal ranges (Andrea, Drechsel \& Starrfield 
1994), except that Iben (1992) found that V842 Cen was the only nova with a He/H 
ratio falling between two groups having solar and half-solar values.

  Recent spectra, however, show peculiarities that are relevant to the present 
state of the primary in V842 Cen. An optical spectrum obtained in 2003 (Schmidtobreick 
et al.~2005) shows a strong blue continuum with weak high ionization lines, e.g. 
C\,IV, probably coming from the nova ejecta but indicative of a hot central 
source, yet accompanied by strong Balmer lines extending to high series components 
and by He\,I emission lines, with only moderate He\,II and 4650 {\AA} Bowen fluorescence 
lines. These are more characteristic of dwarf novae spectra than of old novae. 
Schmidtobreick et al.~(2005) find that the much older nova XX Tau (Nova Tauri 1927, 
$t_3 = 42$ d) has a similar unusual emission line spectrum and suggest that both of 
these novae could be currently in states of low rates of mass transfer ($\dot{M}$).

   Finally, previous high speed photometry of V842 Cen, carried out in 2000 
(Woudt \& Warner 2003: hereafter WW03), showed extreme activity with flares up to 0.25 
mag on time scales $\sim 5$ min but no evident orbital or short period coherent 
brightness modulations, though there were quasi-periodic oscillations (QPOs) on time 
scales $\sim 1000$ s. A parallel was drawn with the light curve of TT Ari, which is a 
high $\dot{M}$ nova-like cataclysmic variable (CV).

    With a view to checking whether V842 Cen had changed its light curve character 
in the 8 years since it was last observed we made an initial exploration in February 
2008, and the surprising result led us to concentrate on it for the remainder 
of that and the following observing run. The optical observations are described 
and analysed in Section 2, X-Ray observations in Section 3, and a discussion 
is given in Section 4.

\section{Optical Observations and Analysis}

Our observations were made with the University of Cape Town's frame transfer CCD 
photometer (O'Donoghue 1995) attached to the 74-in Radcliffe telescope at the 
Sutherland site of the South African Astronomical Observatory. All photometry was 
unfiltered (i.e. in white light), with 6 s integrations, and a white dwarf standard 
star was used to provide an approximate V magnitude scale. The observing runs 
are listed in Table~\ref{v842centab1}.

\begin{table}
 \centering
  \caption{Observing log of photometric observations}
   \begin{tabular}{@{}llccc@{}}
 Run      & Date of obs.          & HJD of first obs. & Length    &  V \\
          & (start of night)      &  (+2450000.0)     & (h)       & (mag) \\[10pt]
 S7791    & 2008 Feb 17 &  4514.54882  &  2.14   &   16.2 \\
 S7795    & 2008 Feb 18 &  4515.45960  &  3.18   &   16.3 \\
 S7798    & 2008 Feb 19 &  4516.45845  &  4.34   &   16.3 \\
 S7803    & 2008 Mar 13 &  4539.41017  &  5.83   &   16.3 \\
 S7806    & 2008 Mar 14 &  4540.39963  &  3.67   &   16.3 \\
 S7809    & 2008 Mar 16 &  4542.39838  &  6.25   &   16.3 \\
 S7811    & 2008 Mar 17 &  4543.43785  &  3.05   &   16.3 \\
 S7813    & 2008 Mar 18 &  4544.40490  &  6.12   &   16.2 \\
\end{tabular}
\label{v842centab1}
\end{table}

\begin{figure}
\centerline{\hbox{\psfig{figure=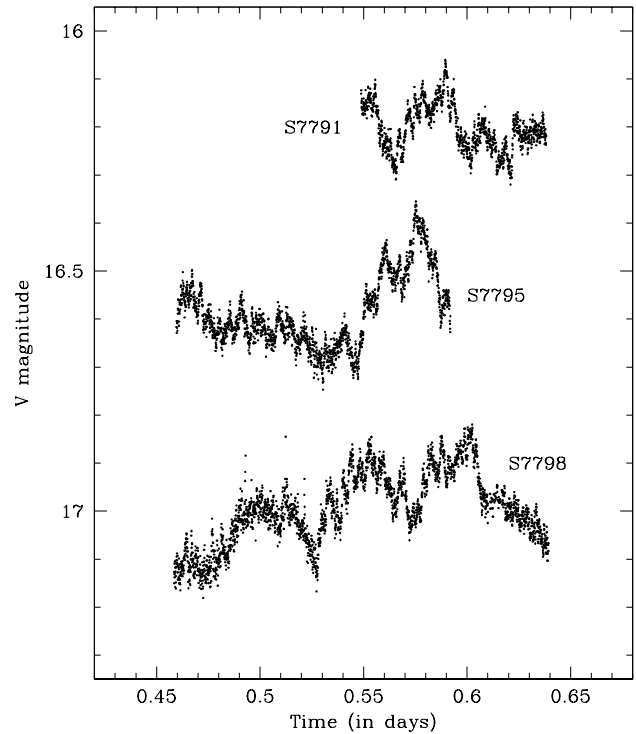,width=8.4cm}}}
  \caption{The light curves of V842 Cen obtained in February 2008. The light curves of S7795 and S7798
have been displaced vertically for display purposes by 0.3 mag and 0.7 mag, respectively.}
 \label{v842cenfig1}
\end{figure}

\begin{figure}
\centerline{\hbox{\psfig{figure=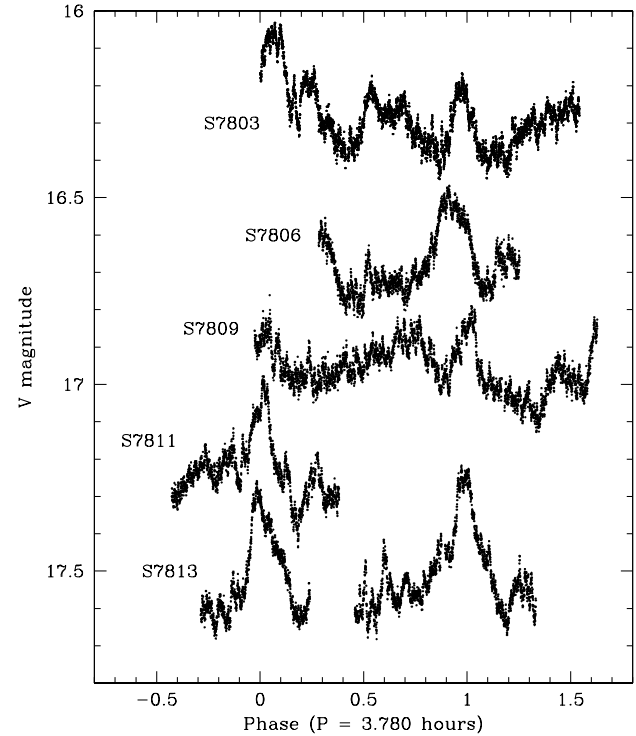,width=8.4cm}}}
  \caption{The light curves of V842 Cen obtained in March 2008. The light curves of S7806, S7809,
S7811 and S7811 have been displaced vertically for display purposes only.}
 \label{v842cenfig2}
\end{figure}

   Fig.~\ref{v842cenfig1} shows the light curves for February 2008 (note that V842 Cen was only 
accessible at the end of the night for a maximum of about 4 hours) and Fig.~\ref{v842cenfig2} 
shows the March 2008 light curves. In the latter we have phased the light curves 
on a period of 3.780 h, for reasons explained below.

\begin{figure}
\centerline{\hbox{\psfig{figure=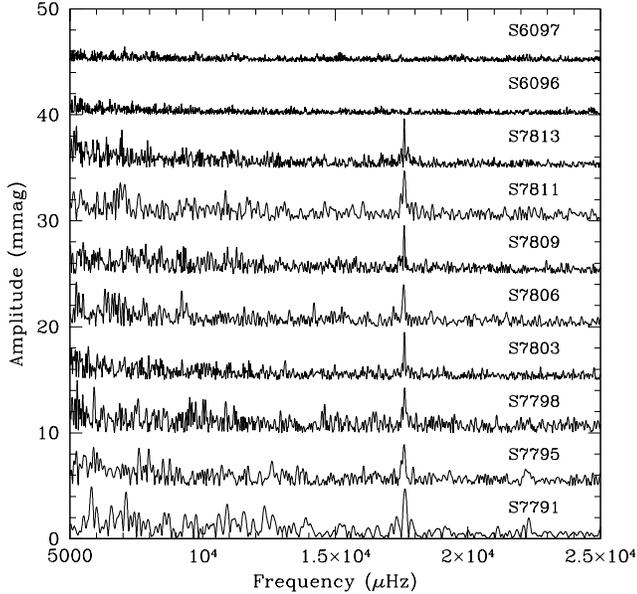,width=8.4cm}}}
  \caption{High frequency fourier transforms of the individual observing runs of V842 Cen.}
 \label{v842cenfig3}
\end{figure}

   Comparison with the 2000 light curves (WW03) shows one immediately obvious 
change -- there are now recurrent peaks of amplitude $\sim 0.3$ mag on a time 
scale $\sim 4$ h. A more subtle addition appears when comparing the high frequency parts of the 
Fourier transforms (FTs) of individual runs in June 2000, February 2008 and March 
2008. As seen in Fig.~\ref{v842cenfig3}, a modulation at $\sim 57$ s has appeared in the interval. 
There is no sign of this in a short run made in March 2002 (WW03).

\subsection{The 57 s brightness oscillation}

\begin{figure}
\centerline{\hbox{\psfig{figure=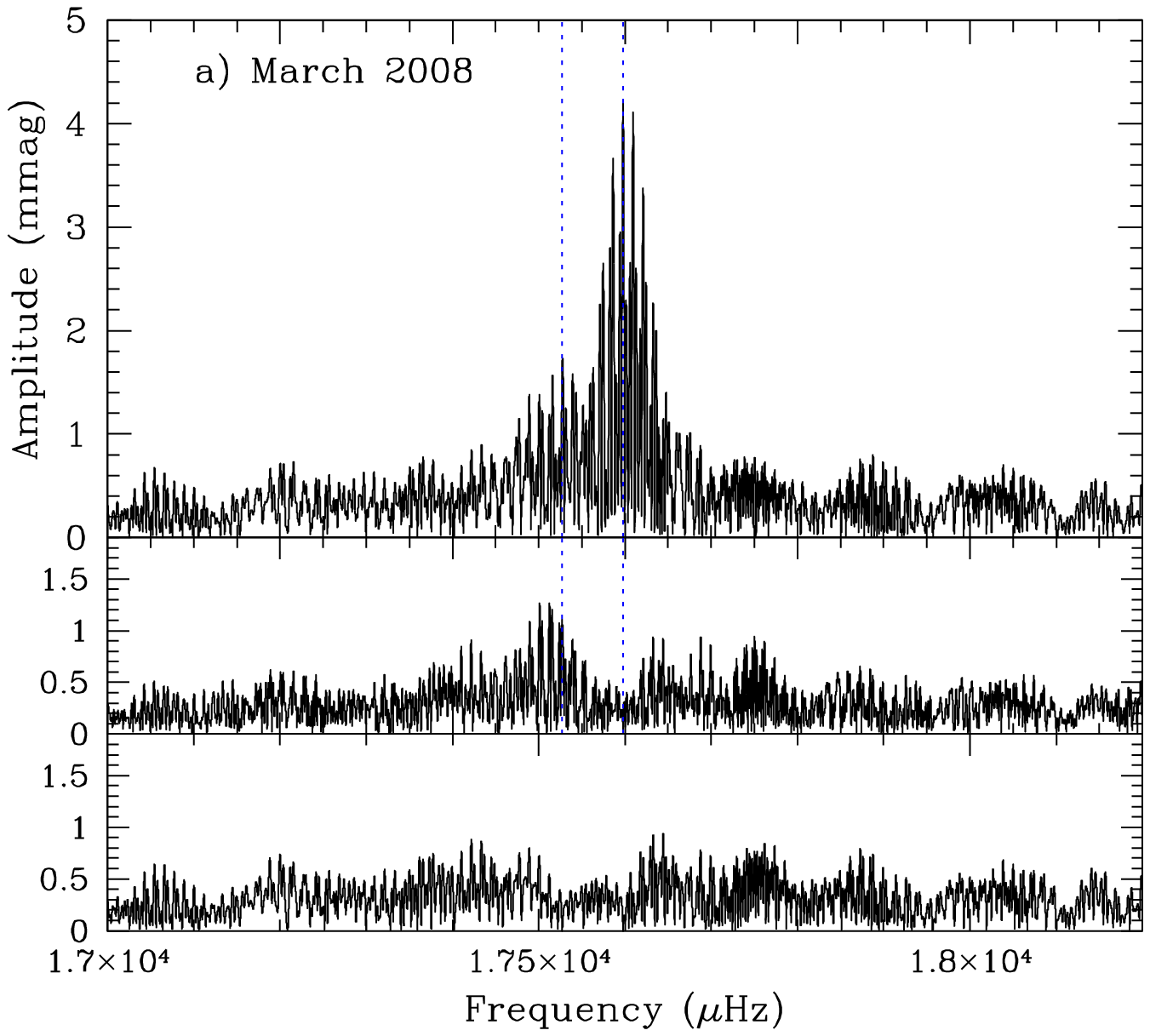,width=8.4cm}}}
\centerline{\hbox{\psfig{figure=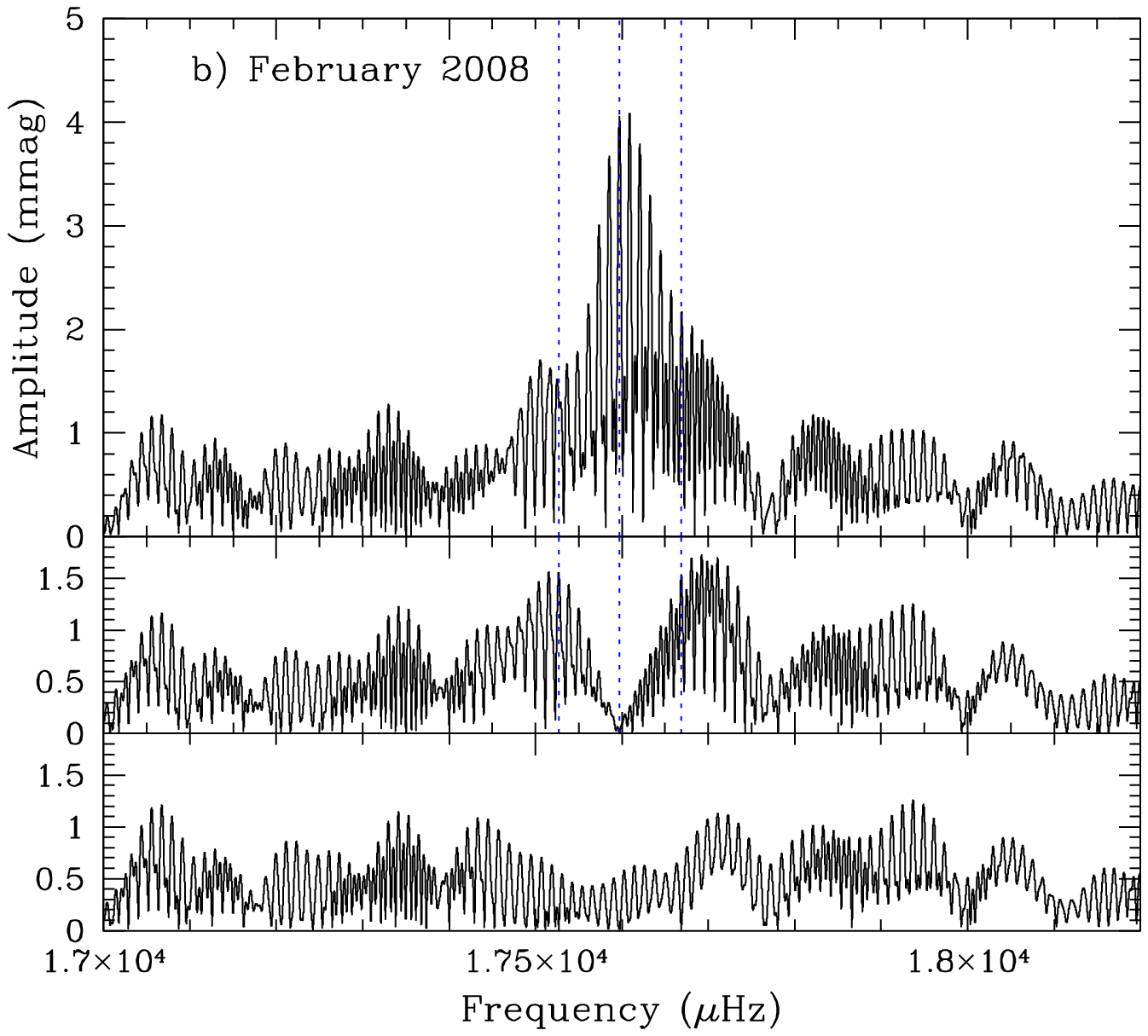,width=8.4cm}}}
  \caption{Fourier transforms in the vicinity of the 57 s oscillations: a.~March 2008 (top diagram), 
b.~February 2008 (bottom diagram). The 57 s oscillation and its orbital sidebands are marked by the dashed 
vertical lines. The middle and lower panels in both diagrams show the Fourier transforms after prewhitening 
with the 57 s oscillation only (middle panels), and the 57 s oscillations plus the identified orbital
sidebands simulaneously (lower panels), respectively. }
 \label{v842cenfig4}
\end{figure}

   We start our analysis by concentrating on the neighbourhood of the $\sim 57$ s signal. 
The FT for the combined March 2008 light curves, which have a five day baseline, is 
shown in Fig.~\ref{v842cenfig4}a (we have omitted run S7813 where the interruption in the light 
curve causes problems in the FT). The dominant feature is the window pattern of the 
data set, centred on 56.825 $\pm$ 0.001 s, with an amplitude of 4.2 mmag (we estimate 
uncertainties from the formal errors of fitting sine curves by least squares). 
Prewhitening with that modulation leaves a signal on the low frequency side (visible 
at the position of the dashed line in Fig.~\ref{v842cenfig4}a) with period 57.054 $\pm$ 0.002 s and amplitude 
1.6 mmag. This is equivalent to a sideband splitting of 70.5 $\pm$ 0.5 $\mu$Hz. Prewhitening with 
both sinusoids simultaneously leaves no significant signal in this region, as seen in the 
lower plot in Fig.~\ref{v842cenfig4}a.

   The FT for the combined February 2008 light curves is shown in Fig.~\ref{v842cenfig4}b. The 
dominant signal over the three day baseline is at 56.828 $\pm$ 0.002 s, and amplitude of 
3.9 mmag. The periods and amplitudes in the two data sets indicate a stable modulation, 
within errors of measurement. As can be seen in the FT, the lower frequency sideband 
is also present, but there is evidence for a longer frequency sideband overlapping the 
principal window pattern. A three sinusoid fit to the light curve gives 57.055 s and 
56.598 s for the two sidebands, both with uncertainty $\pm$ 0.005 s and amplitude 1.7 mmag. 
Prewhitening with these three modulations leaves no significant signal in the region, 
as seen in the lower plot of Fig.~\ref{v842cenfig4}b.

   The frequency difference between the principal signal and the longer frequency 
sideband is 71.6 $\pm$ 2.3 $\mu$Hz, which is within errors the same as the splitting on the 
low frequency side. This arrangement, of equally split sidebands, even with a variation 
of amplitude in one sideband, is the recognizable structure of an intermediate polar (IP). 
Denoting the spin frequency of the white dwarf primary as $\omega$ and the orbital frequency as $\Omega$, 
we have detected the components $\omega$, $\omega - \Omega$, and $\omega + \Omega$, which are characteristic of 
an IP (Warner 1986). The variable amplitude of the one sideband is consistent with 
reprocessed radiation from a rotating source, not simply amplitude 
modulation of a single source.

\begin{figure}
\centerline{\hbox{\psfig{figure=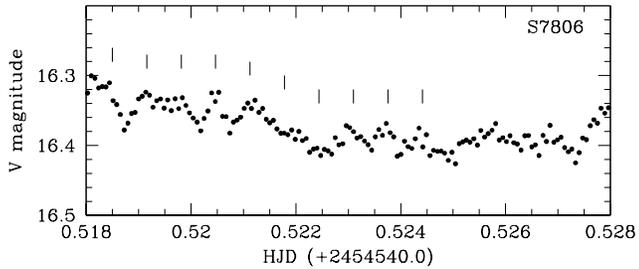,width=8.4cm}}}
  \caption{A section of the light curve of run S7806. The 57 s oscillations are marked by the vertical bars.}
 \label{v842cenfig5}
\end{figure}

   The FTs for individual runs do not resolve the sidebands and as a result the amplitude 
of the 57 s modulation varies on the 70.0 $\mu$Hz, or 3.94 h, time scale. Selecting a 
section where the amplitude is maximal enables us to show the 57 s directly in the light 
curve -- Fig.~\ref{v842cenfig5}.

   In the FTs there are no signs of harmonics or a subharmonic to the 57 s modulations. 
The uncertainties in the 56.825 s signal are just too large to enable the gap between 
the February and March 2008 observations to be bridged without ambiguity, so a more 
accurate period cannot be offered.

\subsection{Orbital and superhump periods}

\begin{figure}
\centerline{\hbox{\psfig{figure=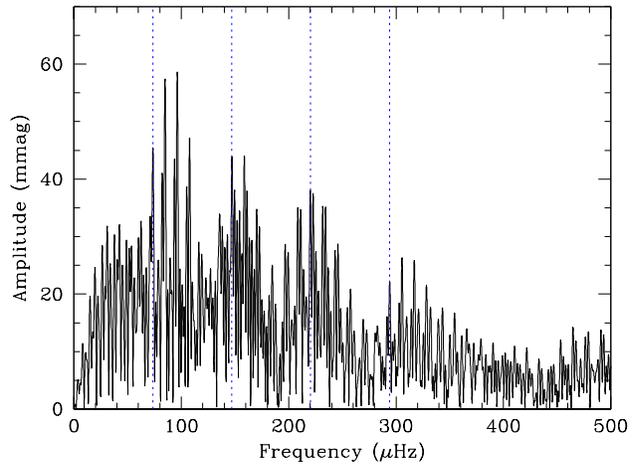,width=8.4cm}}}
  \caption{The low frequency Fourier transform of the March 2008 observations. The 73.46 $\mu$Hz 
frequency and its first three harmonics are marked by the vertical dashed lines.}
 \label{v842cenfig6}
\end{figure}

\begin{figure}
\centerline{\hbox{\psfig{figure=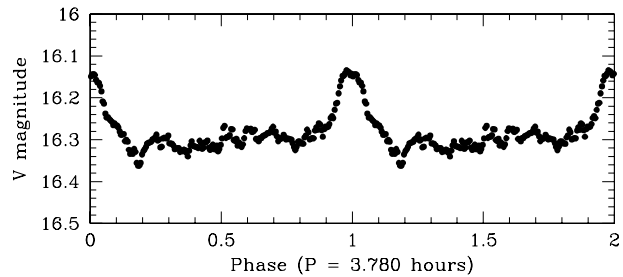,width=8.4cm}}}
  \caption{The average light curve of V842 Cen (March 2008), folded on the 3.780 h superhump period.}
 \label{v842cenfig7}
\end{figure}

   We now turn to the low frequency end of the FTs, seen in Fig.~\ref{v842cenfig6} for the March 
2008 runs (the February data are too sparse to give useful additional information). 
The strongest signal is at 96.23 $\pm$ 0.01 $\mu$Hz, or 2.886 h, with amplitude 60 mmag and no 
detectable harmonics, which we will discuss later. Prewhitening at that frequency 
leaves a strong signal at 73.46 $\pm$ 0.07 $\mu$Hz, or 3.780 h and strong higher harmonics. 
This is the period chosen to phase the light curves in Fig.~\ref{v842cenfig2}. The average light curve 
of the March 2008 data at this period is shown in Fig.~\ref{v842cenfig7}. From the general appearance 
of the light curve -- its narrow and variable peak profiles -- it looks more like a 
superhump modulation than an orbital modulation. The mean amplitude of the peaks is $\sim 0.20$ mag. 
From the discussion in the previous section we would expect that any orbital 
modulation would appear at 70.0 $\pm$ 0.5 $\mu$Hz, equivalent to a period of 3.94 $\pm$ 0.03 h. 
But there is no evidence for this in the FT. But not to despair -- it was already pointed 
out in WW03 that the absence of a $P_{\rm orb}$ signal implies that V842 Cen probably has a low 
orbital inclination. On the other hand, the amplitudes of superhumps are known to be 
independent of inclination (e.g. Warner 1995), so the existence of large amplitude superhumps 
is not a surprise. But what we see is that in V842 Cen they are negative superhumps, with 
a period 4.1\% shorter than the inferred $P_{\rm orb}$. Again we draw attention to the similarities 
between the light curve of V842 Cen and that of TT Ari, which has $P_{\rm orb}$ = 3.30 h and for much of 
the time negative superhumps at a period 3.4\% shorter (Skillman et al.~1998).

\subsection{Quasi-periodic oscillations}

\begin{figure}
\centerline{\hbox{\psfig{figure=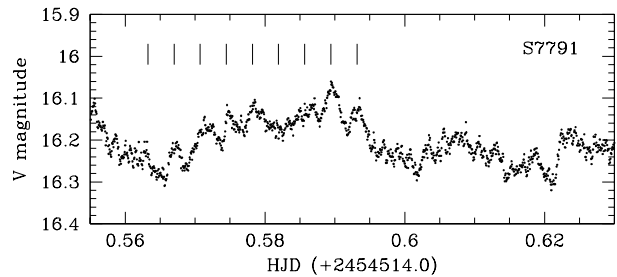,width=8.4cm}}}
  \caption{A section of the light curve of run S7791. The $\sim$ 350 s quasi-periodic oscillations 
are marked by the vertical bars.}
 \label{v842cenfig8}
\end{figure}

We see no sign of any dwarf nova oscillations (DNOs), which might be expected with periods $< 60$ s, 
but the light curves and FTs show occasional presence of QPOs over the range 350 -- 1500 s, 
often with re-occurrence during a particular run.  We show here only an example of a 
train of $\sim 350$ s QPOs in one of the light curves (Fig.~\ref{v842cenfig8}).

\subsection{The 2.89 h modulation}

\begin{figure}
\centerline{\hbox{\psfig{figure=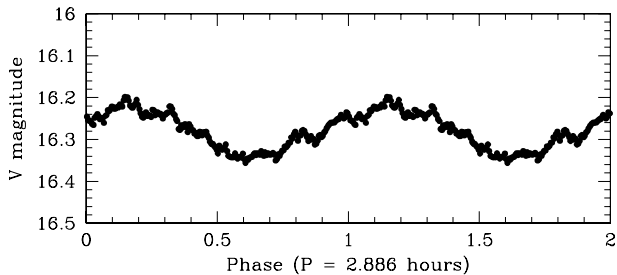,width=8.4cm}}}
  \caption{The average light curve of V842 Cen (March 2008), folded on the 2.886 h period.}
 \label{v842cenfig9}
\end{figure}

The strong signal at 2.886 h is sinusoidal, as seen in the mean light curve of the March 
2008 data: Fig.~\ref{v842cenfig9}. This period bears no obvious relationship to the orbital or 
superhump periods; its ratio to $P_{\rm orb}$ is 1/1.37. Our only current suggestion is that it 
is another example of the mysterious `GW Lib' phenomenon, of which other examples are 
given in table 2 of Woudt, Warner \& Pretorius (2004). There, the periodic signals in 
GW Lib, Aqr1, FS Aur and HS 2331 are listed; the last named has a modulation with a 
period that is also smaller than its $P_{\rm orb}$.

\section{X-ray and UV observations}

The multi-wavelength Swift satellite (Gehrels et al.~2004) observed V842 Cen in July 2008; 
the two twin-snapshot observations are summarised in Table~\ref{v842centab2}. The target 
was clearly seen by the X-ray telescope in Photon Counting mode (Burrows et al.~2005) in 
both observations. The UV-optical telescope (Roming et al.~2005) was operated in the blocked 
filter wheel position for the first observation, and with the uvw1 filter (central wavelength: 
2600 Angstroms) in event mode for the same exposure time as the XRT in the second. 

\begin{table}
 \centering
  \caption{Swift X-ray \& UV observations}
   \begin{tabular}{@{}cccc@{}}
2008 Date of obs. & XRT exp. & XRT count rate & uvw1 \\
(UT)              & (s)      & (c/ks)         & (mag) \\[10pt]
Jul 11 14:03 - 16:05 &  2998  &  $7.4\pm 1.9$  & $-$ \\
Jul 20 10:07 - 13:49 &  2954  &  $7.9\pm 2.0$  & $16.63\pm 0.02$ \\
\end{tabular}
\label{v842centab2}
\end{table}

The data were reduced using version 2.9 of the Swift analysis software and calibration database. 
XRT grade 0-12 $0.3-10$ keV events from V842 Cen were extracted from a $35''$ radius region around 
the source, with the background taken from a surrounding annulus; the count rates corrected for PSF 
losses and bad pixels are given in Table~\ref{v842centab2}. Like the count rates, XRT spectra from 
the two observations were compared and found to be consistent. We used {\sc xpsec} to perform 
unconstrained spectral fits to the total X-ray spectrum, fitting an optically thin plasma emission 
model (mekal) absorbed by cold gas (phabs); having just 35 source counts, we used Cash statistics 
in finding the best fit parameter values and their 90\% confidence errors. These 
are $N_H = (2.2^{+2.6}_{-1.7})\times 10^{21}$ cm$^{-2}$, consistent with the reddening reported 
in Section 1, and $kT = 3.3^{+15.0}_{-1.5}$ keV; the best fit has C-stat $ = 34.5$ for $32$ bins. 
The observed $0.3-10$ keV flux is $2.4\times 10^{-13}$ erg cm$^{-2}$ s$^{-1}$, corresponding to a 
bolometric luminosity at 1 kpc of $5 - 8\times 10^{31}$ erg s$^{-1}$, where the range accounts for 
the uncertanty in the fitted parameters. We used standard Fourier and period-folding techniques, 
but with the small number of X-ray counts, it is not 
surprising that we find no evidence for variability at the periods reported in this work.

The UVOT uvw1 magnitudes were calculated by running the Ftool {\sc uvotmaghist} on the 
two sky images, using the standard source aperture of $5''$. These magnitudes were consistent, 
and the flux-average magnitude is reported in Table~\ref{v842centab2}. Cleaned UVOT event lists 
were created following the standard steps in the UVOT Users' 
Guide\footnote{http://swift.gsfc.nasa.gov/docs/swift/analysis/} and light-curves 
extracted using {\sc uvotevtlc}; they show mild variability possibly like the quieter 
sections of the V-band light-curves of Figs.~1 and 2. We made fast Fourier transforms 
of the 1~s binned UVOT light-curves, but found no evidence of the 57~s period nor of the 
350~s QPO; the two 1505~s and 1445~s light-curves separated by 2.9 hr do not allow a search 
for the longer periods. Folding the two lightcurves together at the 56.825~s period, we used 
sine wave fitting to derive a 90\% confidence upper limit to the modulation amplitude of $<3$\%. 
This can be compared to the 1150-2500 Angstrom continuum DQ Her spin period amplitudes 
of $4.5$\% and $<2$\% measured by Silber et al.~(1996).

\begin{figure}
\centerline{\hbox{\psfig{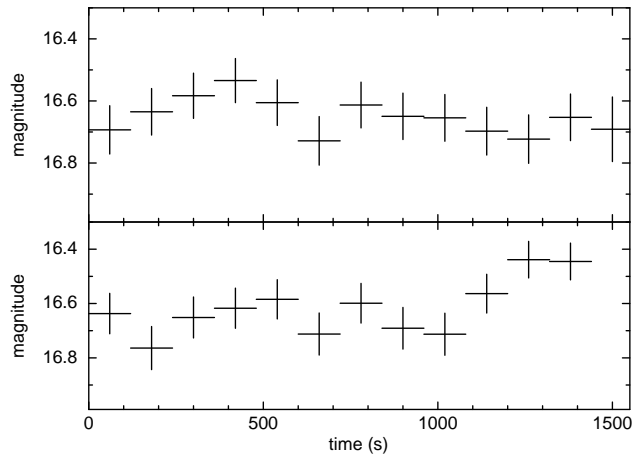}}}
  \caption{Swift UVOT observations of V842 Cen taken in the uvw1 filter on 20 July 2008. 
The top panel starts at UT 10:06:55, the bottom panel at 13:24:55; the bin width is 120s.}
 \label{v842cenfig10}
\end{figure}

\section{Discussion}

A variety of brightness modulations have been found in other nova remnants. 
Of stable oscillations the best known, of course, is that at 71.1 s found in 
DQ Her (Nova Herculis 1934) by Walker (1956). A similar case was later 
found -- the 63.63 s in V533 Her (Nova Herculis 1963), but was only visible 
for a few years (Patterson 1979). More recently, several other nova remnants 
have been found to have stable short periodicities (e.g. Nova Per 1901, 351 s: 
Watson, King \& Osborne 1985; Nova Sct 1975, 258 s: Woudt \& Warner 2003). From 
the essential mono-periodicity (though often with orbital sidebands) of these 
examples, and their similarity to the more slowly rotating IPs (see Chapters 
7 and 8 of Warner (1995)), the periodicities in all such systems are ascribed 
to rotation of magnetic white dwarf primaries.

   V842 Cen has the shortest white dwarf rotation period currently known for 
a nova remnant, and the third shortest known solid body rotation for any CV 
white dwarf -- the other two being the nova-like AE Aqr ($P_{\rm rot}$ = 33.06 s) and 
the dwarf nova WZ Sge ($P_{\rm rot}$ = 27.87 s) (Warner \& Pretorius 2008). There are 
reasons for classifying the shortest period IPs into a separate subclass -- the 
DQ Herculis class (see introduction to Chapter 8 of Warner (1995)), in which case 
V842 Cen is the first far southern hemisphere DQ Her star.

   DNOs, oscillations of lower coherence and short period, common among high 
$\dot{M}$ CVs such as dwarf novae in outburst and in nova-like variables, and 
DNO-related QPOs (for a review see Warner (2004)), have been found in 
only two nova remnants, namely RR Pic (Nova Pictoris 1925), where transient 
oscillations are seen in the range 20 -- 40 s, with preference for $\sim 32$ s 
(Warner 1981; Schoembs \& Stolz 1981) and GK Per, where $\sim 5000$ s oscillations 
are related to the 351 s rotation of the primary (Hellier et al.~2004).  
A current explanation for DNOs is found in magnetic accretion from the 
inner regions of the accretion disc onto a low inertia, slipping equatorial 
accretion belt. If the primary has a field strong enough to prevent the 
formation of a slipping equatorial belt DNOs are not expected, leading to 
mutual exclusion of DNOs and solid body rotational signals (Warner 2004), 
which is as seen in V842 Cen.

    General kilosecond QPOs, although common in nova-likes, are also found in 
only a few novae (e.g. BT Mon and V533 Her) -- Warner (2004). But here again 
there is similarity between V842 Cen and TT Ari -- the latter has QPOs extending 
over the range 900 -- 1500 s (Kim et al.~2008).

The lack of observable modulation in the Swift data from V842 Cen is not constraining. 
However the bolometric X-ray luminosity of $5 - 8\times 10^{31}$ erg s$^{-1}$ is significantly 
lower than the values for the IPs tabulated by Warner (1995) and the average quiescent 
$<L_{\rm Bol}> \sim 3\times 10^{33}$ erg s$^{-1}$ derivable from Ezuka and Ishida (1999) when 
distance is accounted for. We note that DQ Her has a $0.5-2$ keV luminosity of 
$3\times 10^{30}$ erg s$^{-1}$ (Mukai et al.~2003), but in this case the high system 
inclination is believed to block a direct view of the accretion regions. Because 
many IPs have complex X-ray absorption, and we do not have many counts in the X-ray 
spectrum of V842 Cen, it is possible that our simple spectral model underestimates 
its true luminosity (and maximum temperature). Even so, a relatively low accretion 
rate for this IP would appear to be a natural conclusion.

   The spectroscopic evidence that V842 Cen could have a low $\dot{M}$ 
(Section 1, above) is in conflict with the photometric results: a 
precessing disc, as deduced from the presence of superhumps, is a 
characteristic of a high $\dot{M}$ system, especially if the orbital period 
is greater than 3 h (e.g. Murray et al 2000). However, a lessening $\dot{M}$, resulting 
from reduction of irradiation-driven mass loss as the white dwarf cools after 
eruption, would be expected, and this could be the reason for appearance of the 
rotation modulation between 2002 and 2008. Indeed, the almost universal absence 
of DNOs among nova remnants is probably caused by the post-eruption high $\dot{M}$ 
crushing weak magnetospheres down to the surface of the primary. Only strong 
fields can make their presence known in the decades after eruption -- these are 
the four nova remnants that are certain IPs, with another thirteen possibles 
(Mukai 2008).

\section*{Acknowledgments}

We acknowledge Dr Claire Blackman for useful discussions about QPOs. We also acknowledge the referee,
Dr Peter Wheatley, for helpful comments.
PAW's research is supported by the National Research Foundation and the University of 
Cape Town; BW's research is supported by the University of Cape Town; JO and KP acknowledge 
the support of the STFC.

\end{document}